\documentclass[fleqn,usenatbib,useAMS]{mnras}

\usepackage{newtxtext,newtxmath}

\usepackage[T1]{fontenc}
\usepackage{ae,aecompl}

\usepackage{graphicx}	
\usepackage{amsmath}	
\usepackage{amssymb}	
\usepackage[separate-uncertainty = true,multi-part-units=brackets]{siunitx}      
\usepackage{commath}      
\usepackage{physics}  
\usepackage[noabbrev]{cleveref}     

\newcommand*\mean[1]{\langle #1 \rangle}
\newcommand{\torus}{\textsc{torus}} 
\newcommand{\HII}{\ion{H}{II}} 
 
\newcommand{\ue}{\textrm{e}} 

\crefformat{equation}{\eqA equation~\eqB #2#1#3)}
\newcommand{\eqA}{}
\newcommand{\eqB}{(}
\DeclareRobustCommand{\pcref}[1]{%
  \begingroup
  \renewcommand{\eqA}{(}\renewcommand{\eqB}{}%
  \cref{#1}%
  \endgroup
}

\crefname{figure}{Fig.}{Figs.}
\Crefname{figure}{Fig.}{Figs.}
\crefname{table}{Table}{Tables}
\Crefname{equation}{Equation}{Equations}
\newcommand{\solar}{\ensuremath{_{\odot}}} 
\DeclareSIUnit\msol{M\solar{}}
\DeclareSIUnit\lsol{L\solar{}}
\DeclareSIUnit\rsol{R\solar{}}
\DeclareSIUnit\zsol{Z\solar{}}
\DeclareSIUnit\erg{erg}
\DeclareSIUnit\yr{yr}
\DeclareSIUnit\micron{\micro\metre}
\DeclareSIUnit\au{au}
\DeclareSIUnit\pc{pc}
\DeclareSIUnit\sr{sr}
\DeclareSIUnit\str{sr}
\DeclareSIUnit\kms{\kilo\metre\per\s}
\DeclareSIUnit\dyn{dyn}

\title[The growth of \HII{} regions around massive stars]{The growth of \HII{} regions around massive stars: the role of metallicity and dust}

\author[A. A. Ali]{
Ahmad A. Ali\thanks{E-mail: A.Ali2@exeter.ac.uk}
\\
Department of Physics and Astronomy, University of Exeter, Stocker Road, Exeter EX4 4QL, United Kingdom
}

\date{Accepted ???. Received ???; in original form ???}

\pubyear{2020}

\begin{document}
\label{firstpage}
\pagerange{\pageref{firstpage}--\pageref{lastpage}}
\maketitle

\begin{abstract}
Gas metallicity $Z$ and the related dust-to-gas ratio $f_\textrm{d}$ can influence the growth of \HII{} regions via metal line cooling and UV absorption.
We model these effects in star-forming regions containing massive stars. We compute stellar feedback from photoionization and radiation pressure (RP) using Monte Carlo radiative transfer coupled with hydrodynamics, including stellar and diffuse radiation fields. We follow a \SI{e5}{\msol} turbulent cloud with $Z/\si{\zsol} = $ 2, 1, 0.5, 0.1 and $f_\textrm{d} = 0.01 Z/\si{\zsol}$ with a cluster-sink particle method for star formation. The models evolve for at least \SI{1.5}{M\yr} under feedback. 
Lower $Z$ results in higher temperatures and therefore larger \HII{} regions.
For $Z\ge\si{\zsol}$, radiation pressure $P_\textrm{rad}$ can dominate locally over the gas pressure $P_\textrm{gas}$ in the inner half-parsec around sink particles. Globally, the ratio of $P_\textrm{rad}/P_\textrm{gas}$ is around 1 (\SI{2}{\zsol}), 0.3 (\si{\zsol}), 0.1 (\SI{0.5}{\zsol}), and 0.03 (\SI{0.1}{\zsol}). 
In the solar model, excluding RP results in an ionized volume several times smaller than the fiducial model with both mechanisms. Excluding RP \textit{and} UV attenuation by dust results in a \textit{larger} ionized volume than the fiducial case. That is, UV absorption hinders growth more than RP helps it.
The radial expansion velocity of ionized gas reaches $+$\SI{15}{\kms} outwards, while neutral gas has inward velocities for most of the runtime, except for \SI{0.1}{\zsol} which exceeds $+$\SI{4}{\kms}. $Z$ and $f_\textrm{d}$ do not significantly alter the star formation efficiency, rate, or cluster half-mass radius, with the exception of \SI{0.1}{\zsol} due to the earlier expulsion of neutral gas. 

\end{abstract}









\section{Introduction}
Stars form in clusters and associations inside giant molecular clouds (GMCs). In the Milky Way, the star formation efficiency (SFE; the proportion of gas mass converted into stars) is of the order 10 to 30 per cent at the cluster scale and a few per cent at the global GMC scale \citep{lada2003}. 
This inefficiency may be explained by stellar feedback driven primarily by O stars \citep{matzner2002}. These massive stars are sources of energy and momentum for the interstellar medium via mechanisms such as photoionization, radiation pressure, stellar winds, and supernovae \citep[e.g. see the review by][]{dale2015}. In this paper, we focus on the radiative processes.

There remains much uncertainty as to the relative impact of the different feedback mechanisms. One way of assessing this is to compare the pressures associated with each process in a particular star-forming region. This has been done for a variety of \HII{} regions in the Small and Large Magellanic Clouds (SMC and LMC, respectively; \citealt{lopez2011,pellegrini2011,lopez2014,mcleod2019}), in the Galactic Centre \citep{barnes2020}, and more recently in the Galactic disc \citep[][submitted]{olivier2020}. Most regions in these studies show low levels of radiation pressure compared to the thermal pressure from ionized gas, while other regions (typically the smaller ones) show the opposite result. This can also vary as a function of distance from the stellar sources, as in 30 Doradus in the LMC, where the inner region is dominated by radiation pressure and the outer parts by ionization.

One parameter that could modify the effectiveness of feedback is the metallicity, $Z$. The LMC and SMC have sub-solar metallicities with $Z\approx\SI{0.5}{\zsol}$ and \SI{0.2}{\zsol}, respectively \citep{russell1992}. 
Observations show that metallicity varies within the Milky Way, decreasing with distance from the Galactic Centre where $Z\approx\SI{2}{\zsol}$ \citep{deharveng2000}. This will factor into radiative transfer processes.
The dominant cooling sources for ionized gas are collisionally excited forbidden lines from metal ions \citep{osterbrock06}. The dust-to-gas ratio is proportional to metallicity \citep{draine2007}, and radiation pressure is exerted primarily onto dust grains which can be dynamically coupled to the gas. Furthermore, the heating/cooling processes of dust grains can inhibit fragmentation around protostars \citep{krumholz2007a,bate2009,offner2009a}, and scaling this with metallicity can alter the density and temperature structure even in low-mass clusters \citep{bate2019}. When considering feedback from massive stars, changing the metallicity and dust fraction can modify the D-type expansion of \HII{} regions. \citet{haworth2015} investigated this with detailed photoionization modelling in 1D, with a single ionizing source in a uniform-density medium. Including dust grains resulted in the attenuation of ionizing radiation, which made the \HII{} region smaller compared to the model without dust. A higher gas metallicity resulted in a lower electron temperature, decreasing the expansion rate of \HII{} regions. Applying this in 3D to turbulent star-forming regions could provide key information for interpreting the observations which probe different metallicity environments.

Over the last decade and a half, there has been a growing body of work in 3D simulations of photoionization to investigate star formation and gas dynamics \citep[including but not limited to][]{dale2005,mellema2006,peters2010,arthur2011,dale2012,walch2012,colin2013,geen2015,howard2016,gavagnin2017,ali2018,kim2018a,zamora-aviles2019,vandenbroucke2019a,bending2020,fukushima2020,sartorio2021}. Models span a wide range of initial cloud conditions across mass, density, morphology, turbulence, and stellar population. In general, SFEs are reduced by the impact of photoionization compared to hydrodynamics-only runs, with some calculations showing only a modest reduction \citep[e.g.][]{dale2014,howard2017a} while others reach the low values seen in the Milky Way \citep[e.g.][]{geen2017}. 

Radiation pressure is now routinely included in models of massive star formation on the scale of individual cores  \citep{krumholz2009,kuiper2010,harries2017,rosen2019,mignon-risse2020} and implementations are becoming more common for the cluster/GMC scale. For example, radiation pressure has been investigated on its own \citep{skinner2015,tsang2018} and in conjunction with photoionization \citep{howard2016,howard2017a,kim2018a,ali2018,ali2019,fukushima2020}. Overall, radiation pressure appears to be a secondary effect at the GMC scale, except possibly at the highest masses/luminosities/surface densities where it becomes more important \citep{fall2010,howard2018a}. 

However, radiative transfer methods for photoionization are often simplified by prescribing a single temperature for ionized gas, using the on-the-spot approximation, or neglecting dust which absorbs UV radiation and reprocesses it into the IR. Models which include radiation pressure typically only use $\sim$2 frequency bins (e.g. ionizing and non-ionizing photons), use an average dust opacity, and neglect the pressure from either direct (stellar) radiation or indirect (dust-processed) radiation, depending on the regime of interest. This does not accurately capture all the microphysics, as wavelength-dependent dust opacities span many orders of magnitude, and dust-processed radiation may undergo multiple absorption/re-emission/scattering events that can particularly affect high-surface density gas \citep{crocker2018}. Likewise, for photoionization, the presence of diffuse ionizing radiation and dust absorption can change the size and morphology of \HII{} regions \citep{ercolano2011,haworth2012,haworth2015}. Furthermore, capturing the metallicity-dependence of the electron temperature requires computing heating/cooling rates which take into account metal ions. 

In previously published models \citep{ali2018,ali2019}, we included these processes using a detailed Monte Carlo radiative transfer scheme for clouds of \SI{e3}{} and \SI{e4}{\msol} and a single massive star. In this paper, we model a \SI{e5}{\msol} cloud with $Z/\si{\zsol} = $ 0.1, 0.5, 1, and 2, with dust-to-gas ratio scaling linearly with $Z$. We have extended the method to track star formation using cluster-sink particles which form on the fly, accrete gas, and sample stars from an initial mass function. They emit photon packets from across the spectrum which then propagate through gas/dust in the interstellar medium. We detail the numerical methods in \cref{sec:numericalmethods}, go through the results in \cref{sec:results}, discuss them in \cref{sec:discussion}, and conclude in \cref{sec:conclusions}. 
%
%
\section{Numerical methods}
\label{sec:numericalmethods}

\begin{figure}
    \centering
	\includegraphics[width=0.9\columnwidth]{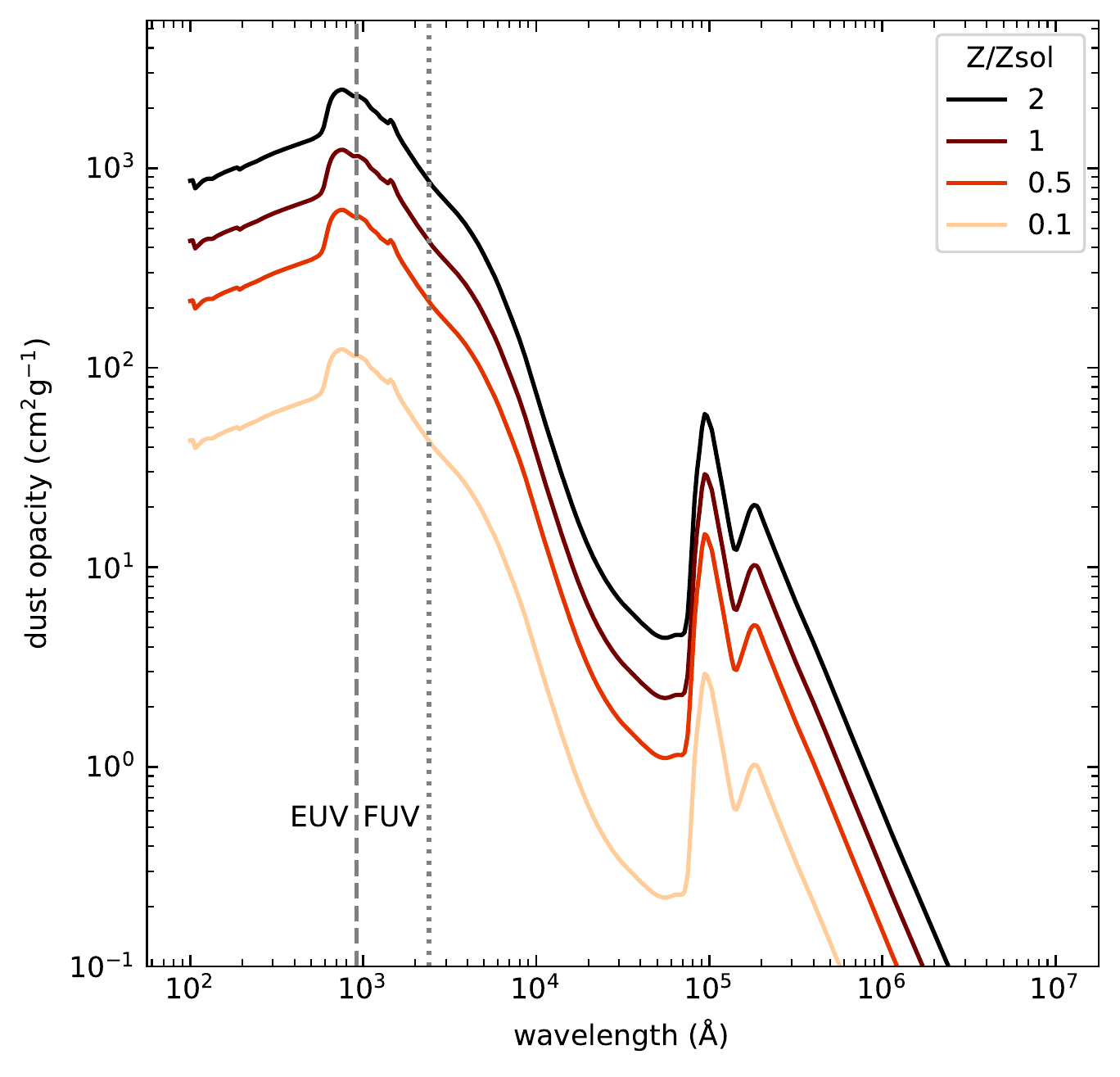}
    \caption{Dust opacities used in \torus{} as a function of wavelength and metallicity $Z$. Also marked are the upper limits of the extreme and far ultraviolet (EUV, FUV, respectively).}
    \label{fig:opacity}
 \end{figure}
\begin{table}
	\centering
	\caption{Abundances at solar metallicity \citep{asplund2009} and ion states used in the photoionization scheme}
	\label{tab:metalabundance}
	\begin{tabular}{lcl} 
		\hline
		Element & $N_X/N_\textrm{H}$ & Ion states \\
		\hline
		Hydrogen & 1 & I--II \\
		Helium & \SI{8.5e-2} & I--III \\
		Carbon & \SI{2.7e-4} & I--IV \\
		Nitrogen & \SI{6.8e-5} & I--III \\
		Oxygen & \SI{4.9e-4} & I--III \\
		Neon & \SI{8.5e-5} & I--III \\
		Sulphur & \SI{1.3e-5} & I--IV \\
		\hline
	\end{tabular}
\end{table} 

We use the \textsc{torus} code, which couples Monte Carlo radiative transfer (MCRT) with hydrodynamics on a 3D grid \citep[for a full description of the code, see][]{harries2019}. We use the same methods as \citet{ali2018} and \citet{ali2019}, with different initial conditions and the addition of a cluster-sink particle approach detailed in \cref{sec:sinkmethods}. As a summary, the hydrodynamics scheme is flux conserving, finite volume, and total variation diminishing. Poisson's equation for self-gravity is calculated using a V-cycling multigrid method with Dirichlet boundary conditions based on a multipole expansion. Gas can flow out of the grid but not in. The hydrodynamics evolves isothermally, with temperatures set by an MCRT calculation at the beginning of each time step. The MCRT scheme is based on the method of \citet{lucy1999}. The source luminosity is split into photon packets which propagate through the medium, undergoing absorption/re-emission/scattering events until they exit the grid. Photoionization equilibrium is computed for the ion states listed in \cref{tab:metalabundance} using the method described by \citet{haworth2012}, with abundances scaling linearly with metallicity. Photon wavelengths are interpolated from 1000 logarithmically spaced bins between \SI{e2}{} and \SI{e7}{\angstrom}. The MCRT scheme therefore follows the diffuse radiation field as well as the stellar radiation field, including ionizing photons re-emitted by gas, and photons reprocessed by dust. The method naturally takes into account shadowed regions, penetration of shadows by diffuse radiation, and radiation hardening. We use \citet{draine1984} silicate grains with an MRN \citep{mathis1977} density distribution,
\begin{equation}
    n(a) \propto a^{-3.5}
\end{equation}
where the grain size $a$ is between 0.005 and \SI{0.25}{\micron}. The dust opacity is plotted in \cref{fig:opacity}. We do not include dust destruction, as on GMC scales this occurs primarily through supernova shocks \citep{jones2004}. We assume the dust is dynamically well-coupled to the gas. Elemental abundances at solar metallicity are taken from \citet{asplund2009} and are listed in \cref{tab:metalabundance}. Thermal balance between photon absorption and thermal emission provides the dust temperature. This is connected to the gas temperature by a collisional heat exchange rate from \citet{hollenbach1979},
\begin{equation}
	\label{eq:gasgraincool}
	\Gamma_\textrm{gas-dust} = 2 f n_\textrm{H} n_\textrm{d} \sigma_\textrm{d} v_\textrm{p}  k_\textrm{B} (T - T_\textrm{d})
\end{equation}
where $T_\textrm{d}$, $n_\textrm{d}$, $\sigma_\textrm{d}$, are the dust temperature, number density, and cross-section respectively, $T$ is the gas temperature, $v_\textrm{p}$ is the thermal speed of protons at $T$, and $f$ is a factor which depends on $T$ and the ionization fraction.
The gas thermal balance uses heating rates from H and He photoionization, cooling from H and He recomination lines, collisionally excited lines from H and metals, and free-free continuum. This takes into account the frequency dependence of the stellar spectra and diffuse radiation field. The momentum-transfer method of \citet{harries2015} provides a value for the radiation pressure, which is added onto the momentum equation in the hydrodynamics step. We do not include magnetic fields in these models.

\subsection{Cluster-sink particles}
\label{sec:sinkmethods}
We use the sink particle algorithm described by \citet{harries2015} which uses formation criteria based on \citet{federrath2010}. Due to the spatial resolution of the calculations we present here, we do not follow the formation of individual stars, so we have extended the method to make sink particles represent clusters or sub-clusters. Sinks are formed above a density threshold of \SI{e4}{\per\cm\cubed}, which corresponds to the density above which stars are observed to form in cores in the Galactic disc (\citealt{lada2003}; \citealt{barnes2019} and references therein; see also \citealt{howard2014} who use this criterion with a similar method in their simulations). 

Before starting the simulations, we pre-tabulate a list of stars by randomly sampling from a \citet{chabrier2003} initial mass function (IMF) up to a total of \SI{e5}{\msol}. This is similar to the method by \citet{geen2018}, but we retain the stars below \SI{8}{\msol} as well. We use the same pre-tabulated IMF for every model. Each sink has a reservoir of mass available for star formation which is replenished by accretion onto the sink. The sink mass gets converted into a stellar population with an efficiency $\epsilon =0.3$, such that the reservoir mass $M_\textrm{reservoir} = \epsilon M_\textrm{sink} - \sum M_{*}$, where $M_{*}$ are the individual stellar masses in that sink. The value of $\epsilon$ is based on observational constraints from \citet{lada2003}; the simulations by \citet{howard2014} use a similar approach with an efficiency of 0.2 per free-fall time. After each hydrodynamical time step, we check which reservoirs are massive enough to take the next star off the list and select one of those at random. The stellar mass is subtracted from the reservoir, and the process continues down the list until no sink reservoirs are massive enough. 

While the position and velocity of each star simply follow the parent sink particle, other stellar properties including mass, radius, and luminosity vary independently with time. These evolve using interpolated values from the MESA Isochrones \& Stellar Tracks \citep[MIST;][]{choi2016} with rotational parameter $v/v_\textrm{crit}=0$ and metallicity [Fe/H] following the gas metallicity set in the initial conditions (\cref{sec:initialconditions}). A sink is allowed to radiate after it is populated with a star more massive than \SI{8}{\msol}, with non-massive stars also contributing to the spectrum. A sink with \textit{only} low-mass stars does not radiate, as the photoionization calculation is more stable without the ionizing flux which is negligible but non-zero. Radiation is emitted from the sink particle as a point source, with the sink's luminosity being the sum of its component stellar luminosities ($L_\textrm{sink} =\sum L_{*}$), and the sink's spectral energy distribution (SED) being an addition of its component stellar SEDs. SEDs for O stars are interpolated from the \textsc{tlusty} OSTAR2002 \citep{lanz2003} grid of models using the relevant gas metallicity, while later-type stars use \citet{kurucz1993} models. In the next hydrodynamical step, the sink particle may continue accreting and the population procedure is repeated. Spectra are only recalculated every ten hydrodynamical time steps, as this reduces the computation time and the hydrodynamical timescale is much shorter than the stellar evolutionary timescale; however, a calculation is forced for sinks which have just been populated.

%
%
\subsection{Initial conditions}
\label{sec:initialconditions}

\begin{table}
	\centering
	\caption{Model parameters}
	\label{tab:models}
	\begin{tabular}{lrrcc} 
		\hline
		Model & $Z$ (\si{\zsol}) & dust/gas & ionization & RP \\
		\hline
		z2   & 2   & 0.02  & \checkmark & \checkmark\\
		z1   & 1   & 0.01  & \checkmark & \checkmark\\
		z0.5 & 0.5 & 0.005 & \checkmark & \checkmark\\
		z0.1 & 0.1 & 0.001 & \checkmark & \checkmark\\
		z1\_norp & 1 & 0.01  & \checkmark & $\times$ \\
		z1\_norp\_nodust & 1 & $10^{-20}$ & \checkmark & $\times$ \\
		hydro  & n/a & n/a & $\times$ & $\times$ \\
		\hline
	\end{tabular}
\end{table}

Our initial condition is a spherical cloud with mass $M=\SI{e5}{\msol}$ and radius $R=\SI{11.9}{\pc}$. These parameters are based on observations of molecular clouds by \citet{roman-duval2010}. The sphere has a uniform-density inner core up to $r=R/2$, then a power-law decrease with $\rho(r) \propto r^{-1.5}$. The density outside the sphere is $0.01 \rho(R)$. The mean mass density and number density of the sphere is \SI{1.05e-21}{\g\per\cm\cubed} and \SI{629}{\per\cm\cubed}, respectively. The free-fall time associated with this average density is $\mean{t_\textrm{ff}}=\SI{2.1}{Myr}$. The mean surface density is $\Sigma=\SI{0.05}{\g\per\cm\squared}=\SI{240}{\msol\per\pc\squared}$. The temperature is initially \SI{10}{K} for both gas and dust. The grid size from end to end is \SI{45.4}{\pc}, giving a linear resolution of \SI{0.18}{\pc} per cell with $256^3$ cells. The grid structure is fixed, uniform, and Cartesian.

We investigate four metallicities, $Z/\si{\zsol} = $ 2, 1, 0.5, 0.1. For each metallicity, we assume a dust-to-gas mass ratio = $0.01 \times Z/\si{\zsol}$ following \citet{draine2007}. For $Z=\si{\zsol}$, we carry out two more models to investigate the effects of radiation pressure and UV absorption: one with dust but no radiation pressure, and one with no dust and no radiation pressure. The models and their labels are summarised in \cref{tab:models}. Metallicity and dust-to-gas ratio remain fixed over the simulation.

We apply a random Gaussian turbulent velocity field over the sphere. This is taken from \citet{bate2002} and has a power spectrum $P(k)\propto k^{-4}$ for wavenumber $k$, such that the kinetic energy equals the gravitational potential energy, i.e. the virial parameter $\alpha_\textrm{vir} \equiv 2 E_\textrm{kin}/E_\textrm{grav} = 2$. We check the criteria for sink particle formation, accretion, and star population as described in \cref{sec:sinkmethods} at every time step (unlike our previous models, \citealt{ali2018} and \citealt{ali2019}, where star particles were placed at 0.75\,$\mean{t_\textrm{ff}}$).

%
%
\section{Results}
\label{sec:results}
\begin{figure*}
    \centering
	\includegraphics[width=0.95\textwidth]{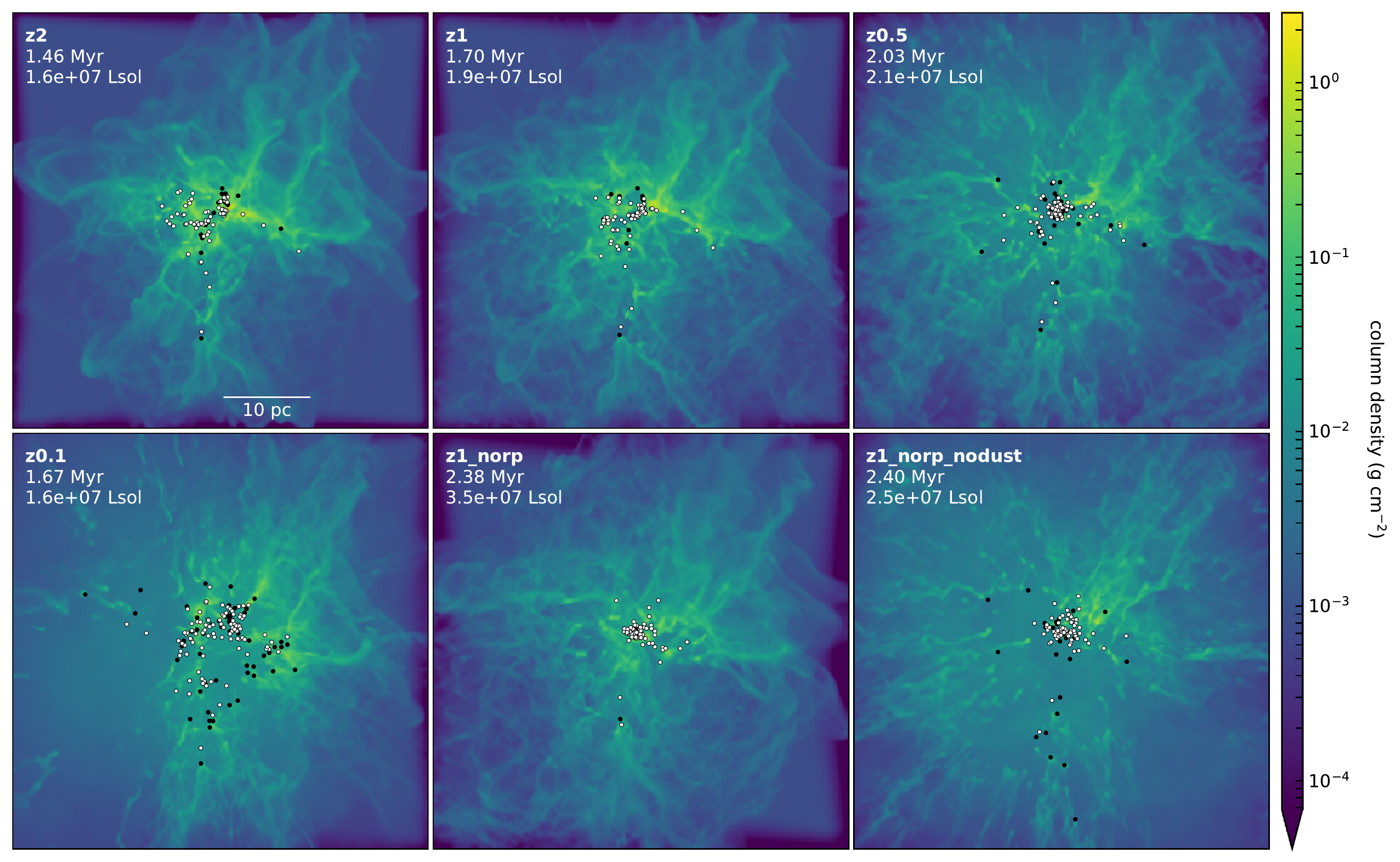}
    \caption{Final snapshots of column density $\Sigma$ integrated along the $z$-axis. White dots show radiating sinks, black dots show non-radiating sinks. Labels show the model name, the time since the formation of the first massive star, and the total bolometric luminosity. The grid is \SI{45.4}{\pc} on each side.}
    \label{fig:finalcolden}
 \end{figure*}
\begin{figure*}
    \centering
	\includegraphics[width=0.95\textwidth]{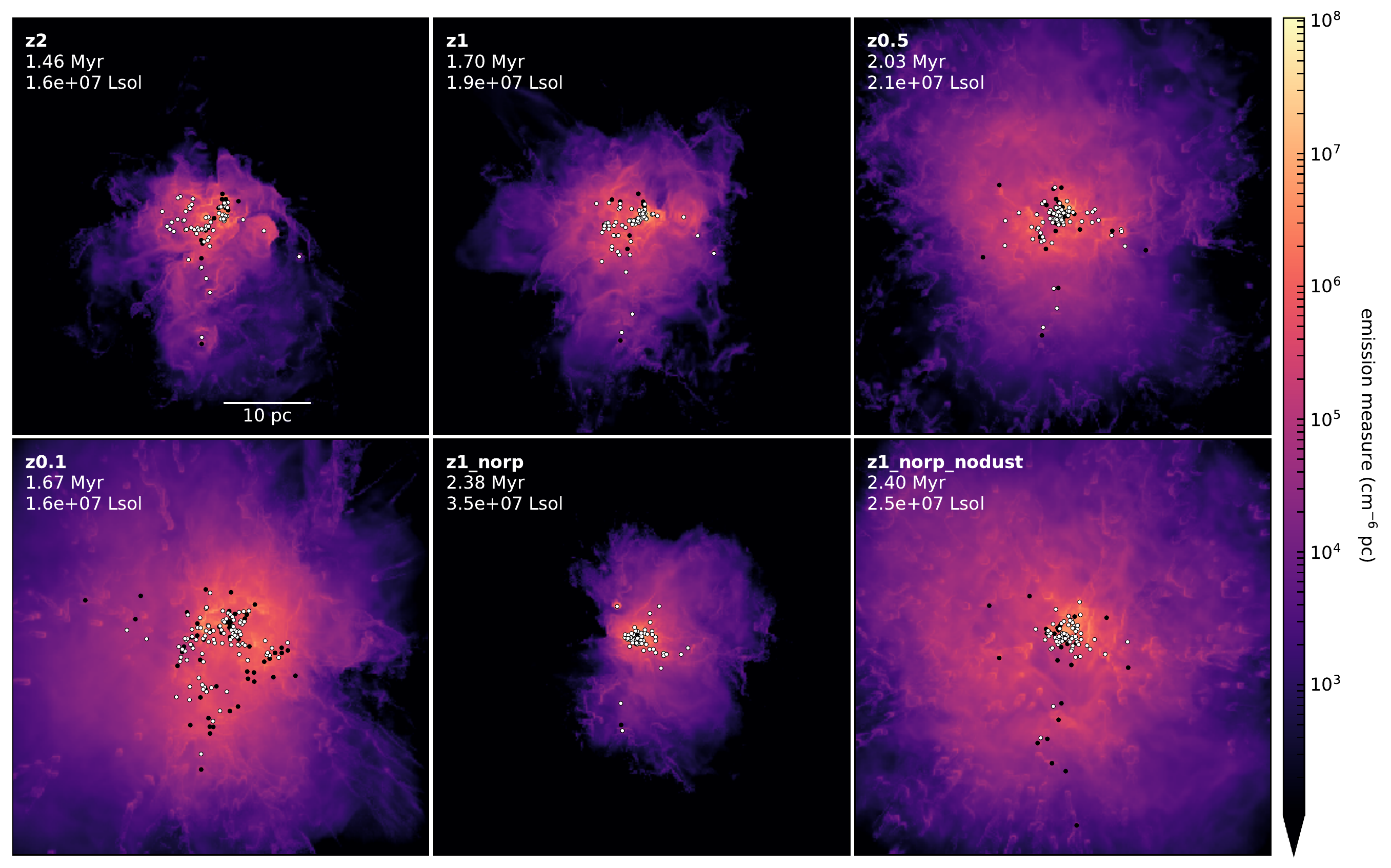}
    \caption{Final snapshots of emission measure $\int n_\textrm{e}^2 \dif{} z$. See also the caption of \cref{fig:finalcolden}.}
    \label{fig:finalemissionmeasure}
 \end{figure*}
 
The spherical cloud evolves under self-gravity and a decaying turbulent velocity field. This creates filaments in which sink particles eventually form, especially in the inner half of the sphere where the density is highest.
The first sink forms after \SI{0.42}{\mega\yr} (0.2\,$\mean{t_\textrm{ff}}$). The first massive star (\SI{19}{\msol}) forms at \SI{0.80}{\mega\yr} (0.38\,$\mean{t_\textrm{ff}}$) -- beyond this point in the text and in all plots, we define this as $t=0$ as this is also when the radiative feedback starts. \HII{} regions are created around sinks which accrete enough mass to form massive stars. As the \HII{} regions expand, they collect gas into shells. Because the density structure is non-uniform, radiation is able to penetrate through lower-density gas, ionizing areas of the cloud further away (including the diffuse environment outside the cloud). Individual \HII{} regions also combine with neighbouring ones. Dense structures resist being ionized, but their star-facing surfaces get photoevaporated. The morphology of the gas and \HII{} regions can be seen in \cref{fig:finalcolden,fig:finalemissionmeasure}, which show the final snapshots of column density $\int \rho \dif{z}$ and emission measure $\int n_\textrm{e}^2 \dif{z}$, where $\rho$ and $n_\textrm{e}$ are the mass volume density and electron number density, respectively. Projected positions of sink particles are shown as circles, with white circles for radiating sinks (those which have massive stars $>$\SI{8}{\msol}) and black circles for non-radiating sinks (no massive stars). The five most massive stars in the z1 model were (in order of formation time),
\SI{78.5}{\msol} (0.56 Myr),
\SI{97.8}{\msol} (0.92 Myr),
\SI{103.5}{\msol} (1.31 Myr),
\SI{88.0}{\msol} (1.42 Myr), and
\SI{115.9}{\msol} (1.69 Myr)
at which point the computation ended.
Simulations were stopped due to small time steps caused by the acceleration from radiation pressure. In \cref{sec:results:hii}, we analyse the growth of the ionized gas as a function of time, and examine the kinematics of both the neutral and ionized components. In \cref{sec:results:sinks}, we calculate star formation rates/efficiencies and cluster properties. Finally in \cref{sec:results:pressures}, we compare the thermal and radiation pressures to determine how each feedback mechanism contributes to the dynamics.

\subsection{\HII{} region expansion}
\label{sec:results:hii}

\begin{figure}
    \centering
	\includegraphics[width=0.9\columnwidth]{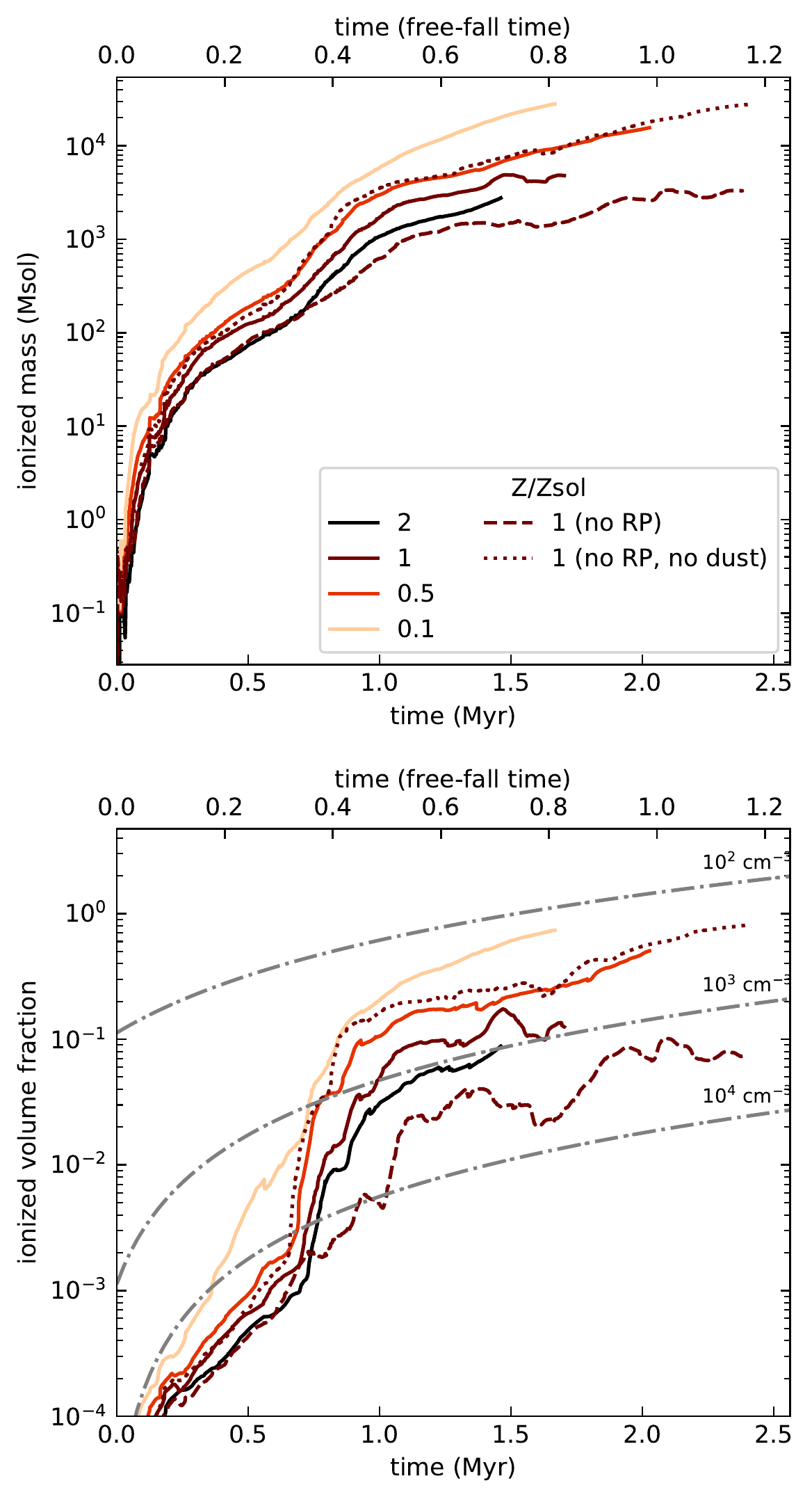}
    \caption{\HII{} region properties - the top panel shows total ionized mass. The bottom panel shows total ionized volume as a fraction of the total grid volume ($45.4^3$\,pc$^3$).  The bottom plot also shows Spitzer expansion profiles (dot-dash lines) for three densities ($n_\textrm{e} = 10^2, 10^3, \SI{e4}{\per\cm\cubed}$, going downwards; see text for more detail). Times in all figures are since the formation of the first massive star and initiation of feedback.}
    \label{fig:plotoftotals_hii}
 \end{figure}

The total ionized mass and volume as a function of time in each model is shown in \cref{fig:plotoftotals_hii}. We count cells as being ionized if the ionization fraction of hydrogen is greater than 0.9. The solid lines show the different metallicities, while the dashed line shows z1\_norp and the dotted line shows z1\_norp\_nodust (see \cref{tab:models} for the parameters and physics in each model).  Lower-metallicity models grow larger \HII{} regions -- a lower cooling rate from metal ions results in a higher temperature, which produces a larger pressure gradient between ionized and neutral gas; furthermore, a lower dust-to-gas ratio means less attenuation of ionizing photons by dust. The mean electron temperatures that we find at each metallicity are \SI{5.6e3}{K} at \SI{2}{\zsol}, \SI{8.2e3}{K} at \SI{1}{\zsol}, \SI{11e3}{K} at \SI{0.5}{\zsol}, and \SI{17e3}{K} at \SI{0.1}{\zsol} -- this is the temperature averaged over the ionized volume then averaged over time. 
The effect of dust absorption and radiation pressure is seen in the difference between z1\_norp\_nodust, z1\_norp, and z1. Of the three, z1\_norp\_nodust grows the largest -- photoionization is the only feedback mechanism, but there is no shielding from dust. Adding dust (z1\_norp) results in the smallest \HII{} region, by an order of magnitude, as now there are dust grains which attenuate the ionizing photons. Enabling radiation pressure (z1) gives a result in the middle, showing that the additional pressure component from dust is not effective enough to counter the stunting caused by absorption. z0.5 has practically the same amount of ionized mass and volume as z1\_norp\_nodust, and z0.1 has the most of all the models. 
For illustrative purposes, we also plot \citet{spitzer1978} expansion profiles as a function of time $t$,
\begin{equation}
	\label{eq:spitzer}
    R_\textrm{I}(t) = R_\textrm{S} \left(1 + \frac{7 c_\textrm{I} t}{4 R_\textrm{S}}\right)^{4/7} ,
\end{equation}
where $R_\textrm{S}$ is the Str\"omgren radius,
\begin{equation}
	\label{eq:stromgren}
    R_\textrm{S} = \left(\frac{3 Q_0}{4 \pi n_\textrm{e}^2 \alpha_\textrm{B}}\right)^{1/3} ,
\end{equation}
for three densities ($n_\textrm{e} = 10^2, 10^3, \SI{e4}{\per\cm\cubed}$), using an ionizing photon production rate $Q_0 = \SI{8e50}{\per\s}$, which is the time-average rate for the longest-running model (z1\_norp\_nodust). The canonical temperature \SI{e4}{K} is used for the ionized sound speed $c_\textrm{I}$, and the case B recombination coefficient $\alpha_\textrm{B}$ = \SI{2.7e-13}{\cm\cubed\per\s}. The Spitzer profiles are not meant to be exactly representative of the modelled \HII{} regions, but they show that the variation between different models is similar to the variation between orders of magnitude in density.

\begin{figure}
    \centering
	\includegraphics[width=0.9\columnwidth]{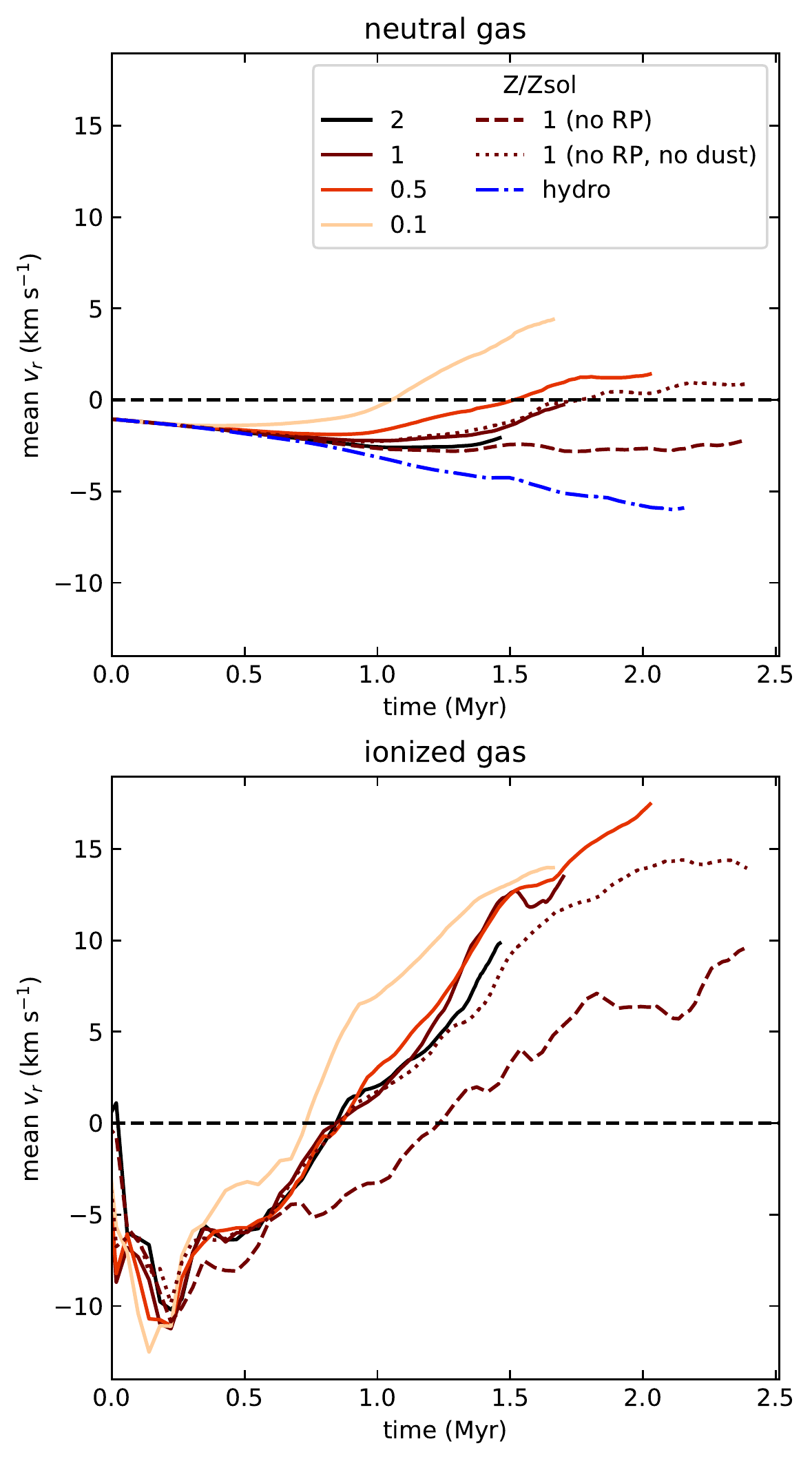}
    \caption{Mass-weighted mean radial velocity in a sphere of radius \SI{10}{pc} around the simulation centre. Positive values indicate outflowing gas, negative values show infalling gas.}
    \label{fig:velocity}
\end{figure}

We calculate the mass-weighted mean velocity in the radial direction,
\begin{equation}
	\label{eq:velocity}
    \mean{v_r} = \frac{\sum_i m_i \mathbfit{v}_i \cdot \hat{\mathbfit{r}}}{\sum_i m_i} 
\end{equation}
summing over cells $i$ with mass $m_i$ and velocity $v_i$ inside a radius of \SI{10}{pc} around the simulation origin. $\hat{\mathbfit{r}}$ is the unit vector pointing away from the origin. This is plotted in \cref{fig:velocity} for each model, separating out the neutral and ionized gas components. In both components, higher velocities belong to lower metallicities. Additionally, the two models without radiation pressure have smaller velocities than the other models, especially z1\_norp (which has UV-attenuating dust and has the smallest \HII{} region). The ionized gas is accelerated radially outwards for the entire duration of every model, with mean velocities reaching in excess of $+\SI{10}{\kilo\m\per\s}$ and still increasing as the models evolve. Velocities can be higher than the ionized sound speed as other dynamical mechanisms are involved, even in the ionization-only models -- the surfaces of dense, neutral material can be photoevaporated, and this newly ionized gas can be accelerated to $\sim$\SI{30}{\kms} \citep[see also][where we show this in a lower-mass cloud with a single massive star]{ali2018}. An improvement on our previous models is that new sink particles form inside dense clumps and ionize them from the inside, creating new compact \HII{} regions. These expand and break out, joining up with \HII{} regions formed by neighbouring sinks.

The neutral gas is infalling for the majority of the runtime for all models except z0.1 and (in the final \SI{0.5}{Myr}) z0.5. Compared to the hydrodynamics-only run, the feedback models have slightly higher velocities (i.e. if gas is flowing inwards, it is doing so more slowly). Furthermore, the gradients eventually turn positive, showing that feedback is accelerating the neutral gas too. The hydrodynamics-only model is the only one which shows the rate of infall continually increasing. $\mean{v_r}$ for z1\_hydro reaches a minimum value of \SI{-5.9}{\kilo\m\per\s}, while the lowest velocity among the feedback models is \SI{-2.8}{\kilo\m\per\s} -- this is reached by z1\_norp, which retains a constant infall velocity over its lifetime. z0.1, the most outflowing model, reaches a velocity of $+\SI{4.4}{\kilo\m\per\s}$. Overall, feedback does affect the kinematics of the neutral gas, providing resistance against infall. However, z0.1 is the only model which exhibits neutral outflow for a significant period of the runtime. In addition to the support given by a higher thermal pressure against gravitational infall, neutral structures can experience a rocket effect when photoevaporated gas pushes off the surface \citep[see also][]{mellema1998,mellema2006,arthur2011}, providing another source of acceleration not present in the hydrodynamics-only model.

\subsection{Star formation}
\label{sec:results:sinks}


\begin{figure}
    \centering
	\includegraphics[width=\columnwidth]{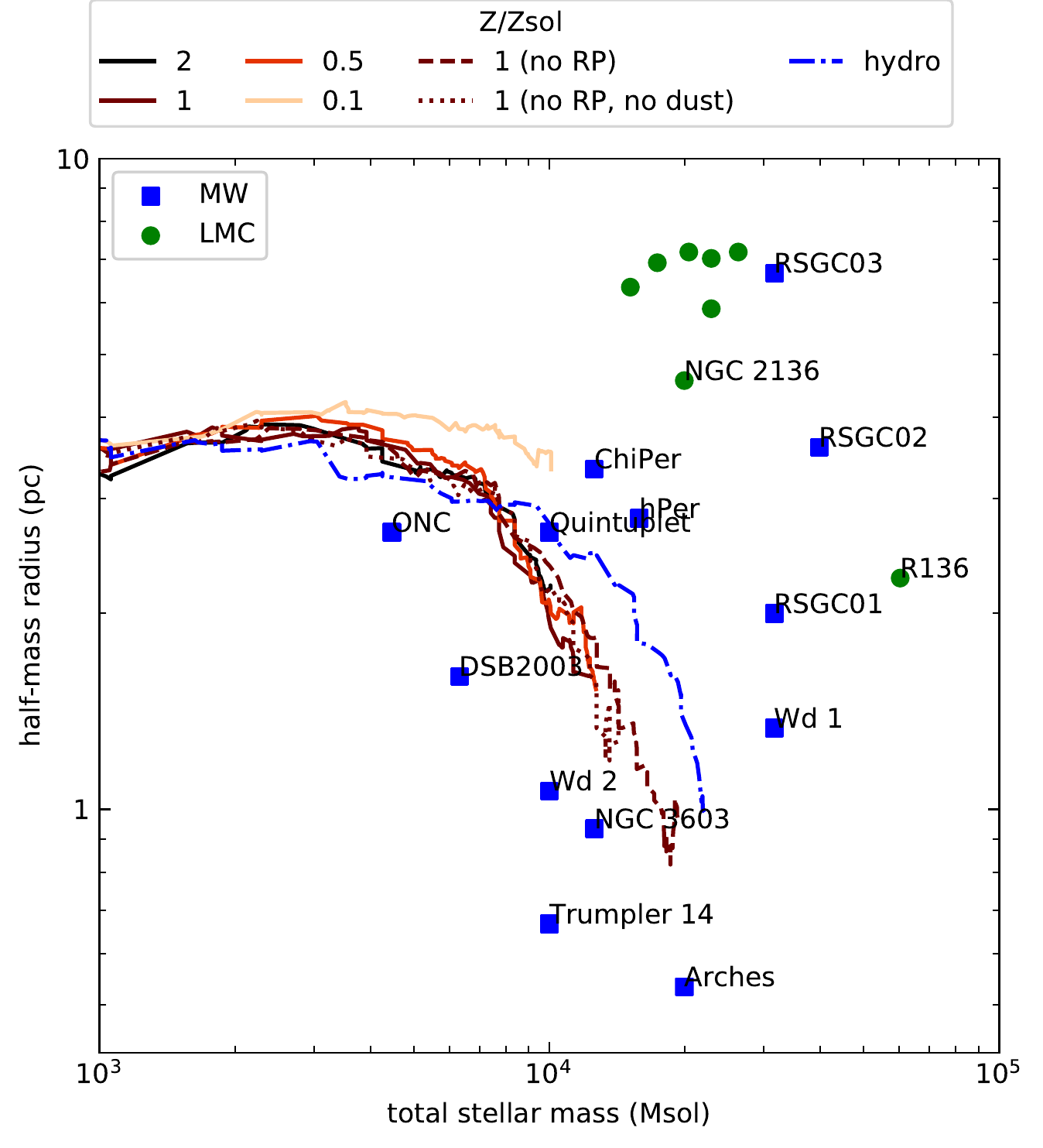}
    \caption{Cluster half-mass radius vs total stellar mass (which increases with time). For comparison, the plot also shows young massive clusters observed in the Milky Way and LMC \citep{portegies-zwart2010}.}
    \label{fig:halfmassradius}
 \end{figure}
Our cluster-sink method allows particles to form and accrete on the fly, which is an improvement on our previous method of placing individual stars manually at one pre-defined moment in time. The drawback is that a particular sink's stars are all lumped together into a point -- instead of having many 1000s of real star particles, we have $\sim$100 `cluster' particles. While this provides a clear computational advantage, it makes it difficult to separate the ensemble of particles into several well-defined, distinct clusters, and to analyse individual clusters by themselves. For this reason, our analysis treats the whole system as one cluster. 

\Cref{fig:halfmassradius} shows the evolution of the cluster half-mass radius $r_\textrm{hm}$ vs total stellar mass for each model. In this case, `cluster' means all sinks which are populated with at least one star, and the relevant mass is the stellar mass not the sink mass (see \cref{sec:results:sinks} for details about the sink particle method). $r_\textrm{hm}$ is the radius from the centre of mass which encloses half the total mass. For comparison, we also show data for young massive clusters (YMCs) in the Milky Way and Large Magellanic Cloud from \citet{portegies-zwart2010}. All the clusters lie between the Milky Way YMCs. With the exception of z0.1, all the models are indistinguishable from one another and become more compact as mass (i.e. time) increases, especially as they approach $\SI{e4}{\msol}$. Similar behaviour is seen by \citet{liow2020} in models of cloud-cloud collisions which do not include feedback, where massive clusters contract once gravity becomes important. The results imply that variations in metallicity, dust absorption, and radiation pressure do not guarantee changes in cluster compactness, even though the \HII{} regions produced by each model are of different sizes and masses (as shown in \cref{fig:plotoftotals_hii}) -- for example, z1\_norp has the smallest \HII{} region, while both z1\_norp\_nodust and z0.5 have an ionized volume which is an order of magnitude larger. That is, a larger \HII{} region does not guarantee a larger cluster. z0.1 is an outlier as its cluster does not contract as much as the other feedback models, which all follow the same relationship as each other. This is because the neutral gas has a higher mean velocity, which is directed radially outwards for z0.1 instead of radially inwards as in the other models (see \cref{fig:velocity}). Since the turnover from infall to outflow occurs relatively early in the evolution (by $t=\SI{1}{\mega\yr}=0.5\,\mean{t_\textrm{ff}}$), this halts the collapse of the gas and hence the sinks -- the turnover happens too late (or not at all) for the other models. The hydrodynamics-only calculation is shifted to the right as it accretes more mass, due to the lack of pressure support from feedback and the accelerating rate of neutral-gas infall.

\begin{figure}
    \centering
	\includegraphics[width=0.9\columnwidth]{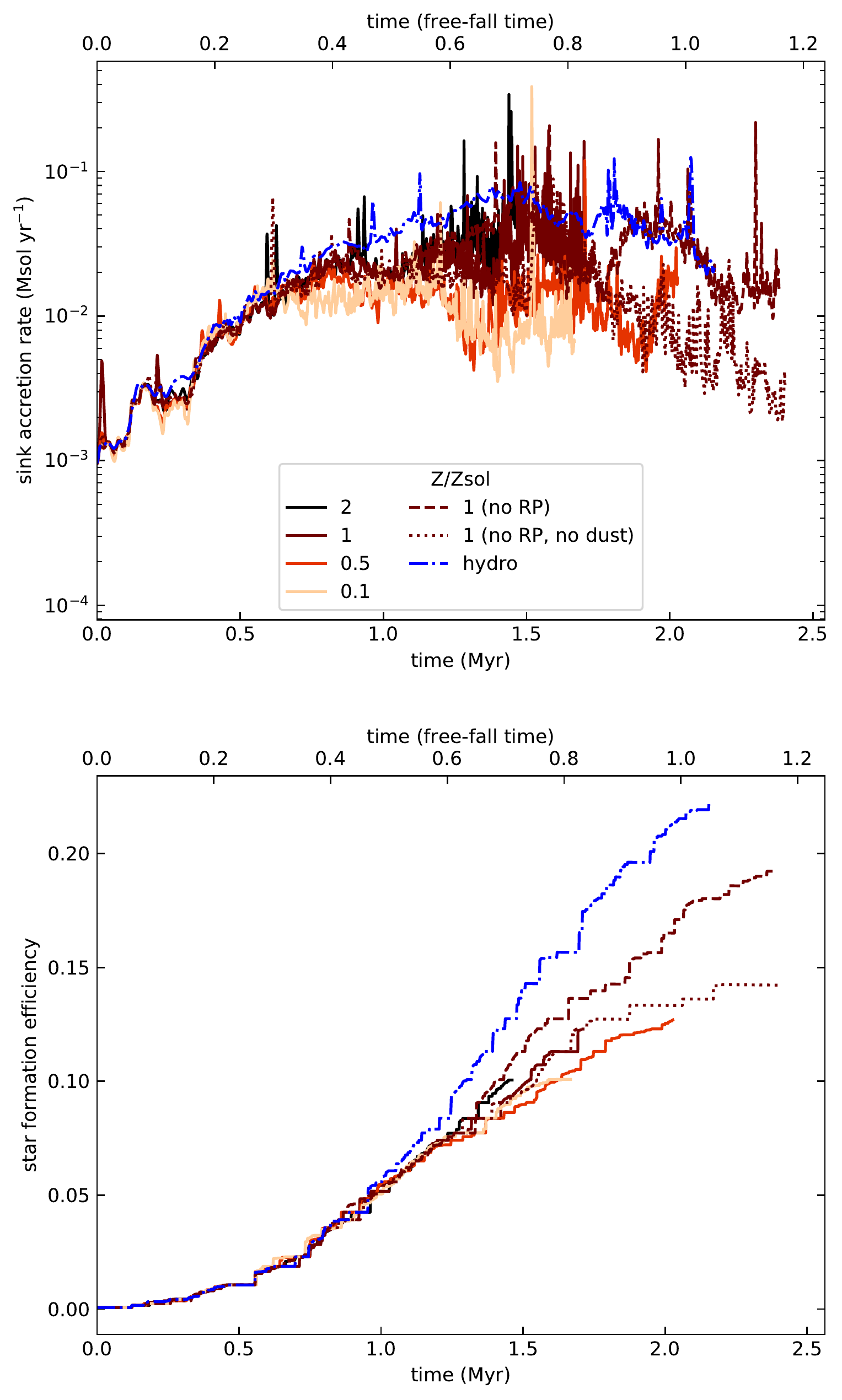}
    \caption{Sink accretion rate (top) and star formation efficiency (bottom).}
    \label{fig:plotoftotals_sfr}
 \end{figure}
 
\cref{fig:plotoftotals_sfr} shows the total accretion rate summed over all sink particles,
\begin{equation}
    \dot{M}_\textrm{sink,tot} = \sum \frac{\Delta M_\textrm{sink}}{\Delta t}
\end{equation}
where $\Delta t$ is the hydrodynamical time step. The result is processed through a $3\sigma$ Gaussian filter to smooth some of the variation and show the trend more clearly. \cref{fig:plotoftotals_sfr} also shows the star formation efficiency over all sink particles relative to the initial cloud mass,
\begin{equation}
    \textrm{SFE} = \frac{\sum M_{*}}{\SI{e5}{\msol}}
\end{equation}
Sink accretion rates lie around \SI{e-2}{\msol\per\yr} for the majority of the runtime, with the hydrodynamics-only run being consistently higher than the feedback models by a factor of a few to 10 at most. The SFE does not exceed 20 per cent within a free-fall time except for the hydrodynamics-only model. The different metallicities cannot be easily distinguished, although they show signs of deviating from each other approximately 0.6\,$\mean{t_\textrm{ff}}$ after the first massive star is formed. Of the models which run the longest, z1\_norp (dashed line) has a higher SFE than z1\_norp\_nodust (dotted line) -- 0.2 vs 0.15. The former has a smaller \HII{} region (\cref{fig:plotoftotals_hii}) and smaller radial expansion velocities both for ionized gas and neutral gas (\cref{fig:velocity}), meaning more mass can funnel into sink particles, creating more stars; the sink accretion rate for this model is higher by about a factor of 10 at late times. Overall the differences between the feedback and hydrodynamics runs are modest, with the hydrodynamics model having approximately 1.5 to 2 times greater SFE. Higher metallicity results in higher accretion rates and SFEs, although we stress the deviation at this stage is marginal.



\subsection{Pressure contributions}
\label{sec:results:pressures}

\begin{figure*}
    \centering
	\includegraphics[width=\textwidth]{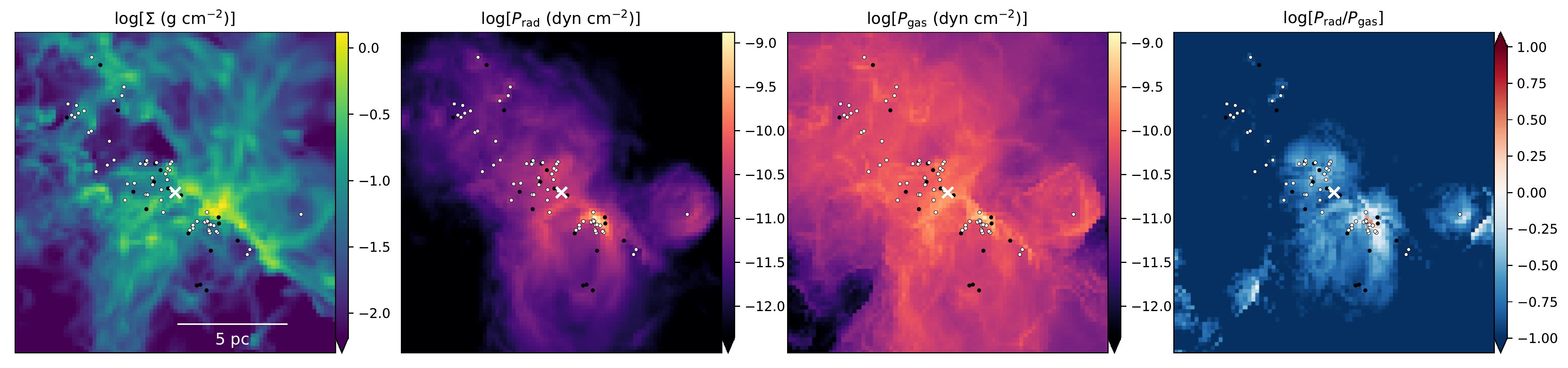}
    \caption{Zoomed-in snapshot showing a radiation pressure-dominated region in the z1 model at $t = \SI{1.47}{\mega\yr}$. $P_\textrm{rad}$ and $P_\textrm{gas}$ are the mean values integrated along the line of sight; the final frame is the ratio of the mean pressures. Images are centred on the cluster centre of mass (shown with an $\times$), with sink particle positions plotted as circles (using the same colour scheme as \cref{fig:finalcolden}). These projections are integrals along the $y$-axis, not the $z$-axis as in \cref{fig:finalcolden,fig:finalemissionmeasure}.}
    \label{fig:pressureprojection}
\end{figure*}

\begin{figure*}
    \centering
	\includegraphics[width=0.85\textwidth]{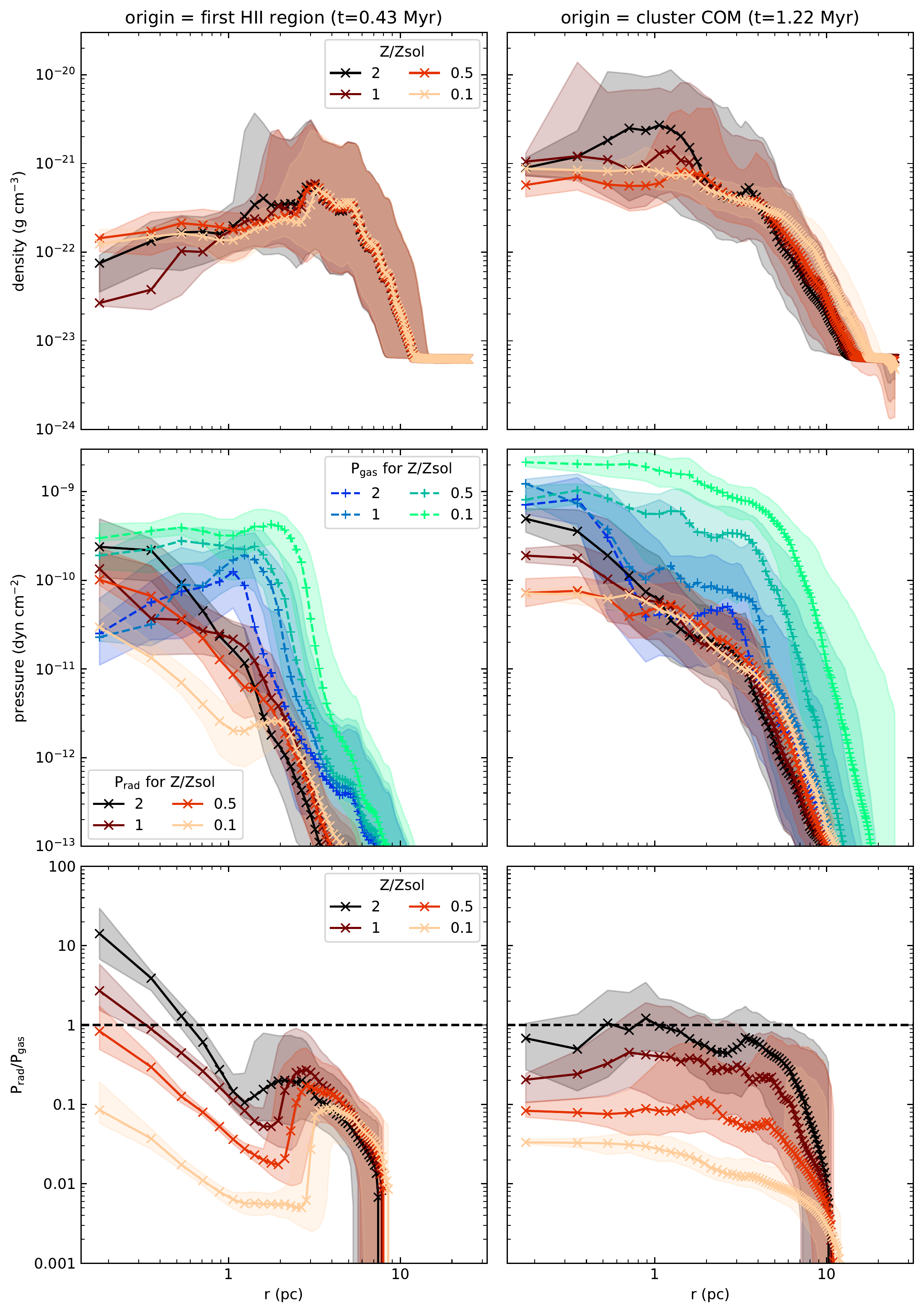}
    \caption{Radial profiles of density, gas thermal pressure $P_\textrm{gas}$, radiation pressure $P_\textrm{rad}$, and the ratio $P_\textrm{rad}$/$P_\textrm{gas}$. Points show the median value in each radial bin. Filled regions show the first and third quartiles in the bin. The origin for the left-hand column is the first massive-star \HII{} region. The right-hand column has the origin at the cluster centre of mass.}
    \label{fig:pressureprofile}
\end{figure*}

In this section, we calculate the radiation pressure as the magnitude of the radiative force per unit area
\begin{equation}
    P_\textrm{rad} = |\mathbfit{f}_\textrm{rad}| \Delta x
\end{equation}
where $\mathbfit{f}_\textrm{rad}$ is the force per unit volume as calculated in the MCRT step, and $\Delta x$ is the cell size. The thermal pressure is given by the ideal gas equation of state 
\begin{equation}
    P_\textrm{gas} = \frac{\rho k_\textrm{B} T}{\mu m_\textrm{H}} 
\end{equation}
where $\rho$ is mass density, $k_\textrm{B}$ is Boltzmann's constant, $T$ is temperature, $\mu$ is the mean molecular weight, and $m_\textrm{H}$ is the hydrogen mass. \cref{fig:pressureprojection} shows a zoomed-in snapshot of the z1 model, plotting the mean $P_\textrm{rad}$ along the line of sight, the mean $P_\textrm{gas}$, and the ratio of the means. This focuses on the inner \SI{7}{pc} around the cluster centre of mass at $t = \SI{1.47}{\mega\yr}$. These projections are analogues of what can be inferred from observations (such as in \citealt{lopez2011,pellegrini2011}; although see caveats in \cref{sec:discussion:pressures}). The region is, on the whole, dominated by the thermal pressure from ionization. Radiation pressure is only significant near the sink particles. For example, the figure shows an area in the \HII{} region which is dominated by radiation pressure, to the bottom right of the centre of mass (the cross symbol). This is where several sink particles are concentrated near a neutral, high-density clump. Additionally, a lone sink particle near the right edge of the frame is growing its own compact \HII{} region and radiation-pressure bubble. To characterise these pressure hotspots and identify trends as a function of radius, we also calculate radial averages of $P_\textrm{gas}$, $P_\textrm{rad}$, and the ratio $P_\textrm{rad}/P_\textrm{gas}$. To do this, we compute each quantity in every cell, bin it according to distance from the origin, then find the median, first quartile, and third quartile in each radial bin. This creates a series of concentric shells around the origin, with average values being found for each shell. These results are shown in \cref{fig:pressureprofile}. 

The origin of the first column of \cref{fig:pressureprofile} is the first \HII{} region at $t=\SI{0.43}{Myr}$. The ionizing sources are the same in each model and have masses of 31.7, 27.8, and \SI{18.7}{\msol}. The \HII{} region is approximately \SI{2}{pc} in radius. In the innermost radial bins, $P_\textrm{rad}/P_\textrm{gas}$ has values around 10 for the z2 model, 3 for z1, 1 for z0.5, and 0.1 for z0.1. That is, radiation pressure dominates over (or is at least comparable to) the gas pressure for all but the lowest metallicity case. This drops off with radius, with all metallicities having $P_\textrm{rad}/P_\textrm{gas} < 1$ by \SI{0.7}{pc}. 

The origin of the second column of \cref{fig:pressureprofile} is the cluster centre of mass at $t=\SI{1.22}{Myr}$. $P_\textrm{rad}/P_\textrm{gas}$ is approximately constant in the inner \SI{5}{pc}, before tailing off at the ionization front. The median is just below 1 for z2, around 0.3 for z1, 0.1 for z0.5, and 0.03 for z0.1. The smaller ratio at low metallicity is a caused by a combination of $P_\textrm{rad}$ being smaller (lower dust-to-gas ratio) and $P_\textrm{gas}$ being higher (higher temperature). 

In summary, radiation pressure can be important at sub-pc scales near sink particles, while ionization becomes the dominant term further out. On the cluster scale, radiation pressure is at best comparable to the gas pressure ($Z \geq \si{\zsol}$), and at worst negligible ($Z = \SI{0.1}{\zsol}$). For the z1 model, the total ionized mass and volume are larger when radiation pressure is enabled compared to when it is switched off (see \cref{fig:plotoftotals_hii}). This can be caused by the high $P_\textrm{rad}$ in the vicinity of sink particles clearing material away with a similar force as the thermal pressure from ionization. However, as explained in \cref{sec:results:hii}, a more effective way of producing a larger \HII{} region is to remove the dust altogether; even though this removes the source of radiation pressure, it means more ionizing photons are absorbed by gas rather than processed away by dust -- feedback is more efficient when photons are ionizing gas instead of imparting momentum to dust.

%
%
\section{Discussion}
\label{sec:discussion}

\subsection{Comparison of pressures with observations}
\label{sec:discussion:pressures}
In this subsection, we compare the pressure contributions with observational studies of the LMC, SMC, and more recently of the Galactic Centre in the Milky Way. These regions have different metallicities, with $Z\approx 2$ in the Galactic Centre, \SI{0.5}{\zsol} in the LMC, and \SI{0.2}{\zsol} in the SMC \citep{russell1992,deharveng2000}. First we consider the different definitions used in the literature. \citet{lopez2011,lopez2014} define the direct radiation pressure as
\begin{equation}
    \label{eq:lopezpressure}
    P_\textrm{dir}(r) = \sum \frac{L_\textrm{bol}}{4 \pi r^2 c}, 
\end{equation}
where the sum is over stars with bolometric luminosity $L_\textrm{bol}$ at a projected distance $r$ away from the point of interest, and $c$ is the speed of light. The volume-averaged value is defined as
\begin{equation}
    \label{eq:lopezpressureavg}
    \mean{P_\textrm{dir}} = \frac{\int P_\textrm{dir}(r) \dif{V}}{\int \dif{V}} = \frac{3 L_\textrm{bol}}{4 \pi R^2 c} .
\end{equation}
where $R$ is the radius of the \HII{} region (or aperture). \citet{mcleod2019} and \citet{barnes2020} use a similar expression, except they assume $L_\textrm{bol} \approx Q_0 \mean{h\nu}$, the rate of ionizing photons emitted by the source population multiplied by the mean SED energy. It is important to note that this does not have a dependence on gas/dust properties (density, opacity, metallicity, etc). Therefore, as noted by the authors, these expressions are strictly a property of the radiation field -- they show the radiative momentum flux at position $r$ given a luminosity which only decreases by geometric dilution. This is delivered as momentum in the gas if radiation streams through an optically thin medium before getting absorbed at the position $r$. Beyond $r$, the deposited momentum would be further reduced by attenuation. Alternatively, radiation could leak out through holes in the density structure instead of interacting with gas at all. Determining this observationally is made more complicated by projecting stars and gas to 2D. 
In short, this $P_\textrm{dir}$ measures the \textit{potential} for a source to provide momentum, not how much momentum is actually deposited in the gas. Without accounting for density, it is possible that $P_\textrm{dir}$ is being overestimated in the observations, particularly for larger radii/optical depths, as $P_\textrm{dir}$ is propagating as $\propto r^{-2}$ instead of $\propto r^{-2} \ue^{-\tau(r)}$. 
\citet{pellegrini2011} use a different method, carrying out a grid of \textsc{cloudy} \citep{ferland1998} photoionization models for varying positions $r$ and densities $n_\textrm{H}$ (keeping $Q_0$ fixed). Therefore instead of \cref{eq:lopezpressure}, they have
\begin{equation}
    \label{eq:pellegrini}
    P_\textrm{dir}(r) = \left(\frac{Q_0}{4 \pi r^2 c n_\textrm{H}}\right) n_\textrm{H} \mean{h \nu} \frac{L_\textrm{bol}}{L_\textrm{ion}}
\end{equation}
where the term in brackets is the best-fitting output of the photoionization models, and $L_\textrm{ion}$ is the ionizing luminosity. This does take into account some of the density and positional information of the source and \HII{} region.

With these caveats, we can compare with the trends found by \citet{lopez2011} \pcref{eq:lopezpressure} and \citet{pellegrini2011} \pcref{eq:pellegrini} who both observed 30 Doradus in the LMC and computed pressure ratios at multiple points inside the region. Both studies found that the direct radiation pressure was greater than or comparable to the ionized gas pressure in the inner region, but was less important further out, by up to an order of magnitude. However, they disagree about the distances involved -- \citeauthor{pellegrini2011} say radiation pressure dominated the inner \SI{10}{pc}, while this is closer to \SI{70}{pc} for \citeauthor{lopez2011}.  The maximum of $P_\textrm{dir}/P_\textrm{gas}$ was around 3 for \citeauthor{pellegrini2011} and 10 for \citeauthor{lopez2011}; the discrepancies may be attributed to differences in the method. Compared with our models, 30 Doradus is an extreme example containing stars with higher luminosities and effective temperatures (thus more energetic spectra). This makes the radiation pressure larger, as $P_\textrm{dir} \propto L_\textrm{bol} \propto T_\textrm{eff}^4$. The ratio $P_\textrm{dir}/P_\textrm{gas}$ is similarly affected, as the ionized gas temperature $T$ does not scale significantly with $T_\textrm{eff}$ (and $P_\textrm{gas} \propto T$).

A later study by \citet{lopez2014} examined 32 \HII{} regions in the LMC and SMC using volume-averaged pressures \pcref{eq:lopezpressureavg}. They found that the ionization pressure was the dominant term, while the direct radiation pressure was up to two orders of magnitude smaller ($\sim$ few $\times$ \SI{e-12}{\dyn\per\cm\squared}). These results were consistent across \HII{} region sizes between 10 and \SI{100}{pc}. \citet{mcleod2019} observed LMC \HII{} regions with higher angular resolution and identified younger, more compact sub-regions (down to \SI{3}{pc}). They reached similar conclusions to \citet{lopez2014}, although values of $\mean{P_\textrm{dir}}$ were even smaller (by a factor of 10) which they attribute to the method being dependent on morphology. \citet{barnes2020} examined \HII{} regions in the super-solar metallicity of the Central Molecular Zone (CMZ) of the Milky Way. The authors note that the CMZ has higher ambient pressures than the LMC/SMC, which would affect the expansion of the \HII{} region and the point where pressure equilibrium between ionized gas and neutral gas is reached (see \citealt{raga2012} for the theoretical description). They used \cref{eq:lopezpressureavg} which is not metallicity-dependent, but the thermal gas pressure (and hence the ratio between them) is, as $P_\textrm{gas} \propto T$. For a sample of \HII{} regions up to a few pc in size, they found a trend with radius showing the direct radiation pressure was dominant up to \SI{0.1}{\pc}, beyond which the ionization pressure became important. The largest ratios were reached on the smallest scales within Sgr B2, with $P_\textrm{dir}/P_\textrm{gas} \approx 10$. 

A final note can be made on the role of the indirect radiation pressure. Although our models do include both components, they can not be disentangled explicitly as we track the \textit{net} momentum change in each cell, not the momentum transferred with each photon interaction event -- in the notation of the observational papers, our $P_\textrm{rad} = P_\textrm{dir} + P_\textrm{IR}$. The observational studies which calculate both components show one of them being much larger than the other. \citet{lopez2014} found $P_\textrm{IR} \sim 0.1  P_\textrm{gas}$ ($\sim$ few $\times$ \SI{e-11}{\dyn\per\cm\squared}), except for two sources where they were comparable, and $P_\textrm{IR} \gg P_\textrm{dir}$. Conversely, \citet{lopez2011} found $P_\textrm{IR} \ll P_\textrm{dir}$. Therefore, it seems reasonable to compare our $P_\textrm{rad}$ to whichever component is dominant for a particular region.  A more analogous comparison could be made by computing synthetic observations from our simulations and using the observational methodology on those outputs. In summary, our work is broadly in agreement with the observational results -- if radiation pressure is the dominant feedback mechanism, this occurs at small scales, while the thermal pressure dominates on larger scales.

\subsection{Comparison with theory}

Our star formation efficiency results are almost the same for each metallicity, although it is possible they may diverge were the models to be evolved for longer. \citet{fukushima2020} did find differences as a function of metallicity in their models of photoionization and radiation pressure at $Z/\si{\zsol} = $ 1, 0.1, and 0.01. They found lower SFEs at lower metallicities, attributing this to the greater disruption at higher temperature. For a similar surface density as our model, the different metallicities showed SFEs of around 20 per cent at \SI{1}{\zsol}, 10 per cent at \SI{0.1}{\zsol}, and 3 per cent at \SI{0.01}{\zsol}. The SFE was set by photoionization -- the fiducial model with both mechanisms was the same as the model with only photoionization, while the model with only radiation pressure had a greater SFE by a factor of 4 (a more drastic variation than we find).

\citet{kim2018a} also found that radiation pressure was secondary to photoionization in their parameter study of GMC mass and surface density. In particular, radiation pressure started to become important above \SI{200}{\msol\per\pc\squared} (for reference, our cloud is \SI{240}{\msol\per\pc\squared}). In general, lower SFEs were reached by photoionization-only models compared to radiation pressure-only runs, and combined-feedback models were slightly lower; around this surface density, the SFEs were around 20 to 40 per cent. It should be noted that \citeauthor{kim2018a} convert 100 per cent of the sink mass to stars, while we limit this to 30 per cent. As in our models, the ratio between the hydrodynamics and feedback SFE was around 0.5.

\citet{howard2017a} also carried out combined photoionization-radiation pressure simulations. They saw SFEs around 16 to 21 per cent in GMCs between \SI{e4} and \SI{e6}{\msol}, which are only slightly higher than the results we find here (albeit their simulations are evolved for a longer fraction of the free-fall time). 
The reductions in the SFE compared with hydrodynamics-only runs were 20 to 50 per cent, which are also exhibited by longer-running models in our set of simulations. Their star formation method and definition of the SFE are similar to ours, perhaps making this a more direct comparison than with \citet{kim2018a}.  
\citet{howard2018a} also applied the two mechanisms in a \SI{e7}{\msol} cloud at $Z/\si{\zsol} = $ 1 and 0.1.  Radiation pressure drove an outflow bubble around the central cluster, but only in the solar metallicity cloud. On those scales, photoionization was insignificant, with the SFE being the same as the hydrodynamics model when radiation pressure was switched off. They found the SFE at \SI{0.1}{\zsol} was higher than the solar model by a factor of 4, which is the opposite trend to \citet{fukushima2020}. However, \citeauthor{howard2018a} only emitted radiation from sink particles once they exceeded \SI{e4}{\msol}, and this could affect the evolution at early times.

The right-most panel of \cref{fig:pressureprojection} shows a region of high radiation pressure where gas is accelerated in the cells around a dense group of sink particles. This creates a cavity in the ionized gas at this location, comparable to the descriptions by \citet{mathews1967} and \citet{draine2011}. It should be noted that dust is dynamically well-coupled to the gas in our numerical scheme. Taking into account separate dust dynamics can result in a dust cavity with variations in dust-to-gas ratio and grain size distribution, as well as shallower gradients in gas density \citep{akimkin2015,akimkin2017,ishiki2018}.

Variation in the initial conditions between simulations clearly adds complexity when trying to make direct comparisons -- the structure of the density and velocity before stars form and feedback initiates will affect cloud evolution.  In addition to mass, average density, and radial dependence of density, different models use different turbulent velocity fields. The effects of this were investigated by \citet{geen2018}, who carried out a suite of models with photoionization feedback, varying the velocity structure and hence final cloud length and filamentary structure. Separately they varied the random sampling of the IMF in the star formation prescription. Resulting SFEs varied between 6 and 23 per cent, showing that a relatively large margin of error can be caused by the model setup and not just the feedback physics. In the context of those results, our work would imply that variations in metallicity are not as important as other initial conditions such as morphology when it comes to setting the star formation efficiency. 

Radial velocities for \citet{kim2018a} were between $+$18 and $+$\SI{26}{\kms} for ionized gas, and between $+$5 and $+$\SI{15}{\kms} for neutral gas. The magnitudes are similar to our results, particularly for the lowest metallicity model, and are consistent with the rocket effect \citep{oort1955,mellema1998,mellema2006,arthur2011}.
The fact that feedback in our clouds must first overcome infall could be due to the initial radial density profile -- \citeauthor{kim2018a} use clouds with uniform density, while ours are centrally concentrated.
The radiation pressure-only model by \citet{tsang2018} shows  infall being stabilised by feedback to a near-constant velocity, while the hydrodynamics run collapses at an accelerating pace. Again, the initial conditions differ, as \citeauthor{tsang2018} use a turbulent box with 100 times more mass and 30 times the density of our cloud, so the infall velocities were also faster -- inside a \SI{1}{pc} radius around the central cluster, this approached $-$\SI{100}{\kms} after a free-fall time in the hydrodynamics run, and around $-$\SI{20}{\kms} in the feedback case. Nonetheless, the qualitative behaviour matches what we find.

Our general results show that photoionization is the dominant feedback mechanism driving the evolution of the cloud, while radiation pressure plays a secondary, local role. This is similar to the dichotomy between photoionization and momentum-only stellar winds as described by \citet{dale2014}, where the momentum injected by winds created cavities around sink particles. Likewise, our SFEs are higher than expected for the GMC scale, which could be attributed to the initial conditions, or to the lack of other feedback mechanisms (such as energy-injecting stellar winds and supernovae which produce hot gas around $\sim \SI{e6}{K}$). While there is still uncertainty about the impact of metallicity on star formation, its role in gas dispersal is clearer -- lower metallicity aids photoionization, while higher metallicity aids radiation pressure.



%
%
\section{Summary and conclusions}
\label{sec:conclusions}
We have modelled a \SI{e5}{\msol} turbulent cloud with different gas metallicities, $Z/\si{\zsol} = $ 2, 1, 0.5, 0.1 and dust-to-gas ratio scaling linearly with $Z$. We included stellar feedback from cluster-sink particles in the form of photoionization and radiation pressure. We also computed thermal balance to consistently calculate electron temperatures at the different metallicities. 
We used a Monte Carlo radiative transfer method which is able to capture the microphysical detail in each feedback mechanism -- photon packets spanning orders of magnitude in wavelength interact with gas and dust grains in the same way as real photons, getting absorbed, re-processed, and scattered. This presents a key advancement compared to previous studies of stellar feedback in GMCs. The method provides robust calculations of ionization pressure, radiation pressure, and hence the ratio between them. As these quantities are used to observationally constrain the relative importance of different feedback mechanisms (e.g. as a function of radius), it is necessary for simulations to provide accurate results. Our key findings are:
\begin{enumerate}
    \item lower-metallicity \HII{} regions have higher temperatures and therefore expand faster. This is because metal forbidden lines are the dominant cooling mechanism for ionized gas. Temperatures range from \SI{5.6e3}{K} at \SI{2}{\zsol} to \SI{17e3}{K} at \SI{0.1}{\zsol}
    
    \item feedback disperses ionized gas and provides support against gravitational infall for neutral gas. Ionized gas is accelerated outwards for the entire duration of each model -- the mass-weighted mean expansion velocity, $\mean{v_r}$, approaches $+$\SI{15}{\kms} by the end of the simulations. Neutral gas has $\mean{v_r}$ directed inwards for most of the evolution. The pure hydrodynamics run is the only model where infall accelerates, reaching $\mean{v_r} = -\SI{5.9}{\kms}$. For the feedback models, the most negative $\mean{v_r}$ is $-$\SI{2.8}{\kms} -- the velocity either stabilises or turns into outflow. This happens in the first Myr ($0.5\,\mean{t_\textrm{ff}}$) at \SI{0.1}{\zsol}, which by the end of its runtime has neutral gas flowing out at $+$\SI{4.4}{\kms}
   
    \item switching off radiation pressure results in a smaller \HII{} region than the fiducial model with both photoionization and radiation pressure (at solar metallicity)
    
    \item however, switching off radiation pressure \textit{and} removing dust results in an \HII{} region \textit{larger} than the fiducial one. This is because UV photons are absorbed by dust grains and reprocessed to lower energies, diluting the ionizing flux. In terms of \HII{} region size, the growth caused by radiation pressure is not enough to completely offset the stunting caused by UV absorption. This highlights the importance of including dust radiative transfer in models of photoionization feedback, whether or not radiation pressure is included
    
    \item we calculate the radial dependence of each pressure component. Radiation pressure dominates over the gas pressure, or is at least comparable to it, in the inner 0.7 parsec around sink particles for $Z\ge\SI{1}{\zsol}$. The maximum ratio of the radiation pressure to gas thermal pressure $P_\textrm{rad}/P_\textrm{gas} \approx 10$ at $Z = \SI{2}{\zsol}$
   
    \item on the global scale, $P_\textrm{rad}/P_\textrm{gas}$ is around 1 at \SI{2}{\zsol}, 0.3 at \si{\zsol}, 0.1 at \SI{0.5}{\zsol}, and 0.03 at \SI{0.1}{\zsol}. 
    
\end{enumerate}

Our results show that although radiation pressure is less important than photoionization on large scales ($\sim\SI{10}{pc}$), it still aids in the growth of \HII{} regions via radiation-pressure hotspots on small scales ($<\SI{1}{pc}$), especially above solar metallicity. This can explain differences in the observations, where \citet{lopez2014} and \citet{mcleod2019} infer negligible radiation pressure in \HII{} regions larger than $\SI{3}{pc}$ at low metallicity, whereas \citet{barnes2020} estimate radiation pressure is dominant on the sub-pc scale at high metallicity. To fully understand how the feedback processes affect the dynamics, it is necessary to observe pressure contributions in young and compact \HII{} regions, in addition to the older, larger \HII{} regions which have until recently been the primary focus of observational studies. Furthermore, this shows the metallicity-dependence of radiation pressure must be taken into account, which is not yet apparent in the literature.

Our models specifically investigate the impact of metallicity and dust in radiative feedback, so we keep the same initial cloud conditions for all models. We do not take into account environmental factors such as variations in external pressure \citep{barnes2020} or galactic-scale forces \citep{rey-raposo2017,bending2020} which can influence the evolution of \HII{} regions and GMCs. We use the same turbulent velocity field for each model and the same ordering of stars which populate cluster-sink particles; varying either can affect cloud morphology and star formation measures \citep{geen2018}. Finally, we neglect other feedback mechanisms such as stellar winds, which may be needed to attain lower star formation efficiencies.

\section*{Acknowledgements}
We thank the referee for an insightful report, and Tim Harries and Clare Dobbs for useful discussions. AAA acknowledges funding from the European Research Council for the Horizon 2020 ERC consolidator grant project ICYBOB, grant number 818940. 
The calculations for this paper were performed on 
DiRAC Data Intensive (DIaL) at the University of Leicester, and the DiRAC@Durham facility managed by the Institute for Computational Cosmology. 
DIaL was funded by BEIS capital funding via STFC capital grants ST/K000373/1 and ST/R002363/1 and STFC DiRAC Operations grant ST/R001014/1.
DiRAC@Durham was funded by BEIS capital funding via STFC capital grants ST/P002293/1, ST/R002371/1 and ST/S002502/1, Durham University and STFC operations grant ST/R000832/1. These form part of the STFC DiRAC HPC Facility (www.dirac.ac.uk). DiRAC is part of the National e-Infrastructure. This paper used NumPy \citep{numpy}, Matplotlib \citep{matplotlib}, Pandas \citep{pandas}, and Astropy \citep{astropy}.

\section*{Data availability}
The data underlying this paper will be shared on reasonable request to the corresponding author.



\bibliographystyle{mnras}
\bibliography{refs}






\bsp	
\label{lastpage}
\end{document}